\documentclass[12pt,a4]{article}

\def\pd{\partial_{\mu}}
\def\pa{\partial}
\def\pu{\partial^{\mu}}

\def\p{\phi}

\def\l{\lambda}
\def\m{\mu}
\def\n{\nu}
\def\e{\epsilon}
\def\s{\sigma}

\def\g{\gamma}

\def\a{\alpha}
\def\b{\beta}
\def\r{\rho}
\def\vp{\varphi}
\def\vt{\vartheta}

\def\F{{\cal F}}

\def\A{{\cal A}}
\def\be{\begin{equation}}
\def\ee{\end{equation}}
\def\ba{\begin{eqnarray}}
\def\ea{\end{eqnarray}}
\def\z{{ }^{*}}
\def\hn{\hat N}
\def\h{\hat}

\def\f0{f_0}
\begin{document}
\thispagestyle{empty}
\vspace{3cm}
\rightline{SINP-2001-37/677}
\vspace{4cm}
\centerline{\large\bf N=2 Supersymmetry and String-Loop Corrected Magnetic
Black Holes }
\vspace{2cm}
\centerline{\bf Mikhail Z. Iofa \footnote{iofa@theory.sinp.msu.ru}}
\centerline{Skobeltsyn Institute of Nuclear Physics}
\centerline{Moscow State University}
\centerline{Moscow 119899, Russia}
\centerline{\today}
\vspace{2cm}

\begin{abstract}
We study string-loop corrections to magnetic black hole. 
Four-dimensional theory is obtained by compactification of the heterotic
string theory on the manifold $K3\times T^2$ or on a suitable orbifold
yielding N=1 supersymmetry in 6D. The resulting 4D theory has N=2
local supersymmetry. Prepotential of this theory  
receives only one-string-loop correction. 
The tree-level gauge couplings are proportional to the
inverse effective string coupling and decrease at small distances from the
center of magnetic black hole, so that loop corrections to the gauge 
couplings are important in this region.  We solve the
system of spinor Killing equations (conditions for the 
supersymmetry variations of the fermions to vanish) and Maxwell equations.
At the string-tree level, we reproduce the magnetic black hole solution
which can be also obtained  by solving the system of the Einstein-Maxwell
equations and the equations of motion for the moduli. String-loop corrections
to the tree-level solution are calculated  
in the first order in string coupling. The resulting 
corrections to the metric and
dilaton are large at small distances from the center of the black hole. 
Possible smearing of the  singularity at the origin  by 
quantum corrections is discussed.
\end{abstract}
\bigskip \bigskip
\pagebreak
\section{Introduction}
There are two ways to obtain classical solutions in supersymmetric 
theories: one can
either solve the equations of motion derived from the effective action 
which, for bosonic fields, are of the
second order in derivatives, or  "spinor Killing equations" resulting
from the requirement that supersymmetry variations of the fermionic fields
vanish. The latter are of the first order in  derivatives. The first
method, in general, provides a larger set of solutions which can include
non-supersymmetric ones. The second way leads to supersymmetric 
solutions with partially broken supersymmetry.

In this paper, we discuss the string-loop corrections to  magnetic 
black-hole solutions, namely, the contributions from the  higher-genera 
topologies  of the string world sheet, by following the second approach. 

The  $4D$  string effective  action is obtained by dimensional reduction 
of $ 6D$, $N=1$ supersymmetric
string effective action on the two-torus.   For this class of
compactifications,  $4D$ theory is N=2 supergravity interacting
with matter. As a concrete example of this construction, we have in view
heterotic string theory compactified on the manifold $K3\times T^2$ or its
suitable orbifold limit, although we do not rely on any  specific properties 
of the model.

Due to N=2 supersymmetry, prepotential of the theory receives only
one-string-loop corrections (from  string world sheets of torus
topology) \cite{afgnt,wikalu}. 
There are  explicit calculations of the loop-corrected
prepotential \cite{afgnt,wikalu,kou,hm}, but for the present study
 only its general structure is important. 

First, solving the string-tree-level "spinor Killing equations" for
gravitino and gaugini, we obtain 
the known spherically-symmetric magnetic black hole solutions \cite{cvyo,bh2}. 
 The tree-level gauge couplings are proportional to the
inverse effective string coupling and decrease at small distances from the
origin, so that the
 loop corrections to the gauge couplings are important in this region. 
 As a technical simplification, we consider tree-level solutions 
with a special relation between the  magnetic charges, in which
case the moduli related to the metric components of the internal two-torus
are constants. 
Next, using the loop-corrected
prepotential, in the first order in string coupling, 
we find the loop-corrected
 gauge couplings, solve the Maxwell equations for the gauge fields  
and  the loop-corrected "spinor Killing equations" for the moduli.

We obtain a family of 
 solutions for the loop corrections to the tree-level metric and dilaton  
 of magnetic black hole which depend on one parameter. 

For a special choice of parameter, in the loop-corrected metric extrapolated to
the region of small distances from the origin, the singularity at the origin
is smeared by string-loop correction.

The set of solutions of the "spinor
Killing equations" is contained in the two-parameter set of solutions 
of the Einstein-Maxwell equations and the equations of motion for the 
moduli derived from the loop-corrected   effective action.  
\section{Heterotic versus $N=2$ pictures}

$4D$ effective string theories obtained by two-torus compactification of
$6D ,\, N=1$ string effective actions share a number of universal 
properties.
The resulting theory is $N=2$ supersymmetric dilatonic supergravity
interacting with matter.  The
bosonic part of the universal sector of this theory written in a 
holomorphic section admitting the prepotential in the standard
form of $N=2$ special geometry  
\cite{dewit,stro,cast,che,afr,cafpr,andr,crap} is
\begin{equation} 
\label{E1}
I_4 = \int d^4 x \sqrt{-g}\left[\frac{1}{2} R + (\bar{N}_{IJ}
\F^{-I}\F^{-J}-
N_{IJ}\F^{+I}\F^{+J} ) + k_{i\bar{j}} \pd z^i \pu \bar{z}^j +\ldots\right] .
\end{equation}
Here $N_{IJ}$ are the gauge coupling constants, 
$$  \F^{\pm }_{\m \n} 
 =\frac{1}{2}(\F_{\m\n} \pm\frac{i}{2} \e_{\m\n\r\l}\F^{\r\l})  
=\frac{1}{2}(\F_{\m\n} \pm i\sqrt{-g}\z\F_{\m\n}).
$$
Here $\z\F_{\m\n} ={1\over2}e_{\m\n\r\l}\F^{\r\l}$, where $e_{\m\n\r\l}$ is
the flat antisymmetric tensor. 

 The moduli $z^i$ are identified below, $k_{i\bar{j}}$ is the K\"{a}hler metric
$$ k_{i\bar{j}}=\frac{\pa^2 K}{\pa z^i \pa \bar{z}^j }. $$

If a holomorphic section of $N=2$ theory admits a prepotential, all the
couplings are defined via this function. In the case of the $N=2$ symmetric
compactification, the prepotential receives
only one-loop corrections and is of the form \cite{afgnt,wikalu,hm}
\begin{equation}
F=-\frac{{X}^1 {X}^2 {X}^3}{{X}^0} -i {{X}^0}^2 h^{(1)}
(-i\frac{{X}^2}{{X}^0},-i\frac{{X}^3}{{X}^0})+\ldots .
\label{E2} 
\end{equation}
where 
\begin{eqnarray}
\frac{{X}^1 }{{X}^0 }=z^1 =iy_1 =i \left(e^{-\p } +ia_1 \right ),
 \nonumber \\
\frac{{X}^2 }{{X}^0 }=z^2 =iy_2 =i \left(e^{\g +\s} +ia_2 \right ),
\nonumber \\
\frac{{X}^3 }{{X}^0 }=z^3 =iy_3 =i \left(e^{\g -\s} +ia_3 \right ),
\label{E3}
\end{eqnarray}
and dots stand for contributions from other moduli. Here and below $I,J
=0,\ldots, 3$ and $i,j =1,2,3$.

The moduli $z^i$ and the vector fields are identified by comparing the
action (\ref{E1}) with that resulting from compactification of the
universal sector of the $6D$  theory
\be
I_6 =\int d^6x \sqrt{-G^{(6)}}e^{-{\Phi}}\left
[R^{(6)} + (\partial {\Phi})^2 - \frac{H^2}{12} \right ]+\ldots.
\label{E4}
\ee
on the two-torus. Here
\be
G^{(6)}=\left(\begin{array}{cc}
 G_{\m\n} + A^m_\m A^n_\n G_{mn}& A^m_\m G_{mn} \\
A^n_\n G_{mn} & G_{mn} 
\end{array}\right),
\ee
where $\m,\n = 0,\dots,3$ and $m,n =1,2$. 
Here $ A^n_\m =G^{nm}G_{m\m}$
The second pair of vector
fields are the components $B_{m\m}$ of the antisymmetric field $B$. 

Dimensional reduction of the action (\ref{E4}) on the two-torus yields the $4D$
action\cite{sen} 
\begin{equation}
\label{F1}
I_4=\int d^4 x\sqrt{-G}e^{-{\p}}\left [R +
(\partial {\p} )^2 - \frac{(H)^2}{12}-{1\over4} \F(LML)\F +\frac{1}{8}Tr
(\partial ML\partial ML) \right ],
\end{equation}
where
\be
\label{B2}
\begin{array}{ll}
M=\left (
\begin{array}{cc}
\,G^{-1} & G^{-1}B\\
-BG^{-1} & \,G
\end{array}
\right ),  \qquad
L=\left (
\begin{array}{ll}
0 & I_2 \\
I_2 & \,0 
\end{array} \right )
\end{array}.
\ee

The metric of the two-torus is parametrized as \cite{duff}
\be
G_{mn}=e^{2\s}\left(\begin{array}{cc}
e^{2\g -2\s} +a_3^2 & -a_3 \\
-a_3 & 1
\end{array}\right)
\label{E5}
\ee
and
$$ \p  = \Phi  -{1\over 2}\ln \det(G_{mn} ).
$$

The dilaton $\phi$ can be split into the sum of the constant part and a
term vanishing at spatial infinity $\phi =\phi_0 +\phi_1$. In string
perturbation theory, higher order
contributions enter with the factor
$e^{{1\over2}\chi {\phi}}$, where $\chi$ is the Euler characteristic of
the string world sheet.
The exponent $e^{\phi_0}\equiv \e $ can be considered as a
string-loop expansion parameter. In the following, we include the factor
$\e$ in string-loop corrections, and use the notation $\phi$ for the
non-constant part of the dilaton.

The moduli (\ref{E3}) are equal to conventional moduli $S, T, U$:
$$(y_1 ,y_2 ,y_3 )=(S=e^{-\p } +ia_1, \,T=\sqrt{G}+iB_{12},
\,U=\frac{(\sqrt{G}+iG_{12})}{G_{22}} ).$$
Here $a_1$ is the axion $\pa_\rho a_1 =
-{H'}^{\m\n\l }e^{-2\p}\sqrt{-g}e_{\m\n\l\rho }$,
the antisymmetric tensor is $B_{mn}=a_2 \e_{mn}$. The tree-level magnetic
black hole solutions we discuss in this paper have $a_i =0$.

The gauge part of the action (\ref{F1}) with $G_{12}=0$ and $B_{12}=0$ is
\be
\label{F2}
-\frac{1}{4}G_{11} (\F^{(1)1})^2-\frac{1}{4}G_{22} (\F^{(1)2})^2
 - \frac{1}{4}G^{11} (\F^{(2)}_1)^2 -\frac{1}{4}G^{22} (\F^{(2)}_2)^2 .
\ee
It is convenient to relabel the vector fields in correspondence with the
moduli with which they form the superfields
\be
A^1_\m =\sqrt{8}\h{\A}^0_\m, \quad B_{1\m} =\sqrt{8}\h{\A}^1_\m ,
\quad A^2_\m =\sqrt{8}\h{\A}^2_\m , \quad B_{2\m} =\sqrt{8}\h{\A}^3_\m .
\label{E6}
\ee
The factor $\sqrt{8}$ appears because of different normalization of the
vector fields in (\ref{E1}) and (\ref{F1}).

Let us turn to the $N=2$ supersymmetric action (\ref{E1}).
In sections which admit the prepotential, the coupling constants in the
action (\ref{E1}) are calculated  using the formula
\be
{N}_{IJ} = \bar{F}_{IJ} +2i \frac{(Im F_{IK}\, X^K ) (Im F_{JL}\, X^L )}
{(X^K \, ImF_{JL}\,X^L )},
\label{E7}
\ee
where $F_I =\pa_{X^I}F, F_{IJ}=\pa^2_{X^I X^J}F$, etc.
Using the loop-corrected prepotential and the formula (\ref{E7}), we
calculate the loop-corrected gauge couplings
${N}_{IJ}$
\ba
\label{E34}
N_{00}& =&iy^3\left(-1+\frac{n}{4y^3 }\right),\quad
N_{01}=-\frac{n+2v}{4y_1 }
-ia_1\frac{y_2 y_3}{y_1 },
 \nonumber \\
N_{02}&=&-\frac{n+2v-2y_2 h y +4y_2 h_2 }{4y_2 }
-ia_2\frac{y_1 y_3}{y_2 }, \nonumber \\
N_{03}&=&-\frac{n+2v-2y_3 h y +4y_3 h_3 }{4y_3 }
-ia_3\frac{y_1 y_2}{y_3 },\nonumber \\
N_{11} &=&-i\frac{y^3}{y_1^2 }\left(1+\frac{n}{4y^3 }\right),\quad
N_{12} =iy_3 \frac{2y_2 h y-n}{4y^3 }+a_3,\quad
N_{13} =iy_2 \frac{2y_3 h y-n}{4y^3 }+a_2,\nonumber \\
N_{22}&=& -i\frac{y^3}{y_2^2 }\left(1-\frac{y_2 h_{23}y_3 }{y^3 }+
\frac{n}{4y^3 }\right),\quad
N_{33}= -i\frac{y^3}{y_3^2 }\left(1-\frac{y_3 h_{23}y_3 }{y^3 }+
\frac{n}{4y^3 }\right)\nonumber \\
N_{23}&= &iy_1 \frac{2yhy -4y_2 h_{23}y_3 -n}{4y^3} +a_1 .
\ea
Here we used the notations: $y^3 = y_1 y_2 y_3,\,\, hy =h_a y_a =h_2 y_2
+h_3 y_3 $, $h_a =\pa_{y_a} h,\, h_{ab}=\pa_{y_a}\pa_{y_b}h$  and
\be
\label{F6}
 v= h-y_a h_a, \qquad n= h- h_a y_a +y_a h_{ab} y_b, \qquad y_2 hy =y_2
h_{ab}y_b .
\ee
In the following, we consider purely real tree-level moduli $y_i$  
which, in particular, is the case for black-hole solutions.
The imaginary parts of the moduli $y_i$
can appear at the first
order in the string coupling constant. Since, the expressions of the first 
order in string coupling are calculated by substituting the 
tree-level moduli, in (\ref{E34})  and below, if it does not 
cause confusion,  
we use notations $y_i$ for the real parts of the moduli $y_i$. 

In sections which do not admit a prepotential (including that which
naturally appears in compactification of the heterotic string action), 
the gauge couplings are calculated by making a symplectic transformation of the
couplings calculated in a section with the prepotential. 
In particular, in the section associated with  compactification 
of the $6D$ heterotic
string on the two-torus,  the string-tree level symplectic transformation
which connects the couplings is \cite{wikalu,cafpr}
\be
O=\left(\begin{array}{cc}
A&B \\
C&D
\end{array}\right),
\label{E8}
\ee
where 
\be
\label{M4}
 A^T C-C^T A=0, \quad B^T D -D^T B =0, \quad A^T D -C^T B=1.
\ee
The gauge couplings are transformed as
\be
\hn=(C+D{N})(A+B{N})^{-1}.
\label{E9}
\ee
Here and below the expressions with hats refer to the section associated
with the heterotic string compactification.
At the one-loop level, we  look for the symplectic transformation in
the form
\ba
\label{M1}
A=diag (1,0,1,1)+\e (a_{ij}),\qquad B=diag (0,1,0,0)+\e (b_{ij}),\nonumber\\ 
C=diag (0,-1,0,0)+\e (c_{ij}),\qquad  D=diag (1,0,1,1)+\e (d_{ij}),
\ea
where $ a\equiv \e (a_{ij}), b\equiv \e (b_{ij}), c\equiv \e (c_{ij})$ and
$d\equiv \e (d_{ij})$ are constant matrices and below the factor $\e$ is not
written explicitly.

Since we perform calculations in
the first order in string coupling constant, we can consider corrections to
the matrices of symplectic transformation and corrections to the gauge 
couplings due to the one-loop term in the prepotential independently. 
From the relations (\ref{E9}) we have
\be
\label{M2}
\h{N}_{(0)IJ}+\h{N}_{(1)IJ}=(c+d\,N_{(0)})(A +B\,N)^{-1}_{(0)}-
\h{N}_{(0)}(a+b\,N_{(0)})(A +B\,N)^{-1}_{(0)}, 
\ee
where the subscripts $(0)$ and $(1)$ refer to the tree-level and first-order
terms correspondingly.

In the heterotic string compactification, the loop corrections to the gauge
couplings appear with the factor $\e e^\p$. The couplings $N_{00}, \,N_{22}$
and $N_{33}$ 
are proportional to $e^{-\p}$, whereas $N_{11}$ contains the factor
$e^{\p}$. Thus, admissible  structures of the one-loop
corrections to the tree-level
gauge couplings $\hn_{IJ}$ in the section connected with 
the heterotic string compactification can be of the form $\frac{c}{N_{00}},
\frac{c}{N_{22}}, \frac{c}{N_{33}}$ and $c N_{11}$, where $c$ is one
of the entries $a,b,c$ and $d$.
Examining the relations (\ref{M2}), we find that 
admissible non-zero matrix elements are $ c_{ij}$ with $c_{1i}=c_{i1}=0$ 
and $d_{11}$.

Calculating the couplings in the holomorphic section associated with the
heterotic string compactification, we have
\be
\h{N}_{IJ}=\left(\begin{array}{cccc}
N_{00}+ c_{00}
-{N_{01}^2 \over N_{11}} & {N_{01}\over N_{11}}
& N_{02}+c_{02}-{N_{01} N_{12}\over N_{11}} &
N_{03}+c_{03}-{N_{01}N_{13}\over N_{11}}\\
{N_{10} \over N_{11}} & -\frac{1}{N_{11}} +d_{11} &
\frac{N_{12}}{N_{11}} &
\frac{N_{13}}{N_{11}}\\
N_{20}+c_{20} -{N_{21} N_{10} \over N_{11}} &\frac{N_{21}}{N_{11}}
&N_{22}+c_{22} -{N_{21}^2 \over N_{11}} &
N_{23}+c_{23} -{N_{21}N_{13} \over N_{11}}\\
N_{30}+c_{30}-{N_{31}N_{10}\over N_{11}} & \frac{N_{31}}{N_{11}} &
N_{32}+c_{23} -{N_{31}N_{12} \over N_{11}}& N_{33} +c_{33} -{N_{31}^2 \over
N_{11}}
\end{array}\right).
\label{D7}
\ee

The terms of the form
$\frac{N_{K1} N_{1J}}{N_{11}}$ are of the next order in  string coupling.

The field equations and  Bianchi identities for the gauge field strengths   
are
\ba
\label{E12}
\pd \left( \sqrt{-g} Im \,G^{-\,\m\n }_I \right) =0 \nonumber \\
\pd \left( \sqrt{-g} Im \, \F^{-J\,\m\n}\right) =0,
\ea
where $G^{-\,\m\n }_I=\bar{N}_{IJ}F^{-J\,\m\n}$. 
Eqs. (\ref{E12}) and  are invariant under the symplectic 
transformations (\ref{E8}) of general form. 

From the symplectic transformation of the field strengths
\be
\label{E15}
\left(\begin{array}{c}\h{\F}^- \\ \h{G}^-\end{array}\right)
=O\left(\begin{array}{c} \F^- \\  G^-\end{array}\right),
\ee
with the matrices (\ref{M1}) 
 we obtain the relations between the field strengths
$$ \h{\F}^{-0} = {\F}^{-0} , \qquad \h{\F}^{-2} = {\F}^{-2 }, \qquad \h{\F}^{-3} = 
{\F}^{-3}
$$
\ba
\label{E16}
\h{G}^-_0& = &G^-_0 +c_{00}\F^{-0} +c_{02}\F^{-2} +c_{03}\F^{-3},\,\,
\h{G}^-_1 = -\F^-_1 +d_{11}G^{-1},\nonumber \\
 \h{G}^-_2& = &G^-_2 +c_{20}\F^{-0} +c_{22}\F^{-2} +c_{23}\F^{-3}, \,\,
\h{G}^-_3 = G^-_3 +c_{30}\F^{-0} +c_{32}\F^{-2} +c_{33}\F^{-3}.
\ea
Substituting in the relation $\h{G}^{- }_I=\bar{\hn}_{IJ}\h{F}^{-J}$ 
the  couplings (\ref{D7}) and relations between the field strengths
(\ref{E16}) (any $I=0,1,2,3 $ can be used ), we obtain the relation
\be
\label{M5}
{\F}^{-1 } =-\frac{\bar{N}_{10}}{\bar{N}_{11}}\h{\F}^{-0}
+\frac{1}{\bar{N}_{11}}\h{\F}^{-1}
-\frac{\bar{N}_{12}}{\bar{N}_{11}}\h{\F}^{-2} -
\frac{\bar{N}_{13}}{\bar{N}_{11}}\h{\F}^{-3} 
\ee
which does not contain arbitrary constants $d_{11}$ and $c_{ij}$.

The K\"{a}hler potential is invariant under symplectic transformations and its
part which depends on the moduli $y_i$ is given by
\be
K=-\ln[(y_1 +\bar{y}_1 +V)(y_2 +\bar{y}_2)(y_3 +\bar{y}_3)],
\label{E10}
\ee
where the Green-Schwarz function $V$ is \cite{kou,wikalu}
\be
V(y_2,\bar{y}_2,y_3,\bar{y}_3) = \frac{Re\,h^{(1)} -Re\,y_2 Re\,\pa_{y_2 }
 h^{(1)}-Re\,y_3 Re\,\pa_{y_3 } h^{(1)}}{Re\,y_2 \,Re\,y_3}.
\label{E11}
\ee
\section{Supersymmetry transformations}
To write the supersymmetry transformations, one introduces the expressions
(for example, \cite{cast,che,afr,cafpr,andr})
\ba
\label{E18}
S_{\m\n}& =& X^I Im\,N_{IJ} \F^{-J}_{\m\n}, \\\nonumber
T^-_{\m\n}& =&2ie^{K/2} X^I Im\,N_{IJ} \F^{-J}_{\m\n}
\ea
and
\be
\label{E19}
G^{-i}_{\m\n} =-k^{i\bar{j}}\bar{f}^I_{\bar{j}} Im\,N_{IJ} \F^{-J}_{\m\n}.
\ee
Here $k^{i\bar{j}}$ is the inverse K\"{a}hler metric, and 
$$f^I_i = (\pa_i +\frac{1}{2} \pa_i K ) e^{K/2}X^I  .
$$

Supersymmetry transformations of the chiral gravitino $\psi_{\a\m }$ and
gaugino $\l^{i\a}$ are (for example, \cite{cast,che,afr,andr})
\ba
\delta \psi_{\a\m} =D_\m \e_\a + T^-_{\m\n}\g^\n \e_{\a\b}\e^\b, \\
\delta \l^i_\a = i\g^\m \pd z^i \e^\a +
G^{-i}_{\m\n}\g^\m\g^\n\e^{\a\b}\e_\b,
\label{E20}
\ea
where
$$
D_\m \e_\a = (\pd -{1\over 4}w^{\h{a} \h{b}}_\m \g_{\h{a}} \g_{\h{b}} + 
{i\over 2} Q_\m )\e_\a .
$$
Here $w_\m^{\h{a} \h{b}}$ is the spin and  $Q_\m$ is the K\"{a}hler 
connection. Here
$\h{a},\h{b},...$
are the tangent space indices, $a, b, ... $ are the curved space indices.

Requiring that supersymmetry variations of spinors vanish, we obtain a
system of supersymmetric Killing equations for the moduli. 
We look for a solution of this system with 
supersymmetry parameter satisfying the relation $\e^\a = 
\g_{\h{0}} \e^{\a\b} \e_\b $. 
The $\m =0$ component of equation $\delta \psi_{\a0}=0$   takes the form
\be
\label{E22}
({1\over 2} w_0^{\h{a} \h{b}} \g_{\h{0}}  \g_{\h{b} } \g_{\h{0}} - 
T^-_{0n}  e^{\h{b}n }\g_{\h{b}} )\e_{\a\b } \e^\b =0.
\ee
Here $e^{\m \h{b} }$ is the inverse vielbein. 

In this paper we are interested in static spherically-symmetric solutions of
the field equations. The metric is
\be
\label{E17}
ds^2 = -e^{2U} dt^2 + e^{-2U} dx^i dx_i.
\ee
The only non-vanishing components
of the spin connection $w_0^{\h{a} \h{b}}$ are $w_0^{\h{0} \h{b}} = 
{1\over 2}\pa_b e^{2U}$. 
The vielbein $e_\m^{\h{b}} $ is $e_\m^{\h{b}} =\delta_\m^b e^{U}$. 
 To have a nontrivial solution for
the supersymmetry parameter, we must require that
\be
\label{E23}
{1\over 2} w_0^{\hat{0}\hat{n}} - e^{U} T^-_{0n}  =0.
\ee

Using the relations
$$G^-_{mn} =i \e _{mnp0}G^{-p0}
$$
and
$$ G^-_{\m\n}\g^\m \g^\n \e_\a = 4G^-_{0n} \g^0 \g^n \e_\a
$$
valid for any self-dual tensor and chiral spinor, the condition 
 the gaugini supersymmetry transformation to vanish is written as
\be
\label{E24}
(i\g^n \pa_n z^i \g^{\h{0}} + 4G^{-i}_{0n} \g^0 \g^n)\e^{\a\b}\e_\b =0.
\ee

There is a nontrivial solution provided
\be
\label{E25}
i\pa_n z^i +4e^{-U}G^{-i}_{0n} =0.
\ee
The factor $e^{-U}$ is due to the relation $\g_{\h{0}} =-\g^{\h{0}}
=-e^{\h{0}}_0 \g^0 = -e^U \g^0 $. Convoluting the equation (\ref{E25}) 
with the functions $f^I_i$
and using the relation of special $N=2$ geometry
$$
k^{i\bar{j}}f^I_i \bar{f}^J_{\bar{j}} =-{1\over2}(Im N)^{IJ} - e^K \bar{X}^I X^J,
$$
 it is obtained in the form (cf. \cite{bgl,fre})
\be
\label{E26}
if^I_i \pa_n z^i + 4e^{-U}\left({1\over 2}\F^{-I}_{0n}+ e^K \bar{X}^I
(X^J\,Im\,N_{JL}\,\F^{-L}_{0n})\right) =0.
\ee
Combining the Eqs.(\ref{E23}) and (\ref{E26}) it is possible to present the
latter as (cf. \cite{wi,mo})
\be
\label{Ea26}
e^{-K/2}\left[e^U \pa_i (e^{K}X^I )\pa_n z^i - (e^{K}\bar{X}^I ) \pa_n e^U \right]
=2i\F^{-I}_{0n} 
\ee
\section{Solution of spinor Killing equations for magnetic black hole}
Next, we solve the combined system of the equations for the gauge field
strengths and the moduli. First, we  look for a string-tree-level solution 
with the metric in the form
(\ref{E17}), two magnetic fields $\h{\F}^0_{\m\n}$ and $\h{\F}^1_{\m\n}$ 
and purely real moduli $y_i$ (\ref{E3}). This means, that we consider
configurations with diagonal metrics $G_{mn}$, vanishing tensor $B_{mn}$
and vanishing axion $a_1$. Such configurations appear as solutions of the 
"chiral null models" \cite{cvyo,bh2}. In the next section we shall solve the
equations including the string-loop corrections.   

Solving the system of Maxwell equations and  Bianchi identities in the
section associated with the heterotic string compactification, we have
\be
\label{E28}
\h{\F}^{-0}_{0n}  =P^{0} {i\over 2}e^{2U} \frac{x^n}{r^3}, \qquad
\h{\F}^{-1}_{0n}  =P^{1} {i\over 2}e^{2U} \frac{x^n}{r^3}.
\ee
\be
\label{E29}
K=-\ln\, 8y_1 y_2 y_3 .
 \ee
The tree-level expression for the symplectic-invariant 
 combination $T^-_{0n}$ (\ref{E18}) which enters the gravitini 
supersymmetry transformations is
\ba
T^{-}_{0n}&=& 
2i e^{K/2}(Im N_{00}\F^{-0}_{0n} + iy_1 Im N_{11}\F^{-1}_{0n})\nonumber \\
&=& 
2i e^{K/2}(-y_1 y_2 y_3 {\h{\F}}^{-0}_{0n} -y_1 {\h{\F}}^{-1}_{0n})
=\left(\frac{y_1 y_2 y_3 }{8}\right)^{1/2}
\left(P^0 +\frac{P^1 }{y_2 y_3 }\right)e^{2U}
 \frac{x^n}{r^3}.
\ea

 The gravitini Killing equation (\ref{E23}) takes the form
\be
\label{E30}
{1\over 4}\pa_n e^{2U} - \left( \frac{y_1 y_2 y_3}{8}\right)^{1/2}
e^{3U} ( P^0 +\frac{P^1}{y_2 y_3} )\frac{x^n}{r^3} =0 .
 \ee

The tree-level gaugini equations (\ref{E26}) written in the section with the
prepotential are
\ba
\label{F3}
I&=&0:\qquad \frac{ie^{K/2}}{2}\pa_n \ln y_1 y_2 y_3 -4e^{-U}\left(\frac{1}{2}
\F^{-0}_{0n} +e^K S_{0n}\right)=0,\nonumber \\
I&=&1:\qquad \frac{y_1 e^{K/2}}{2}\pa_n \ln \frac{y_2 y_3}{y_1} +4e^{-U}
\left(\frac{\h{\F}^{-1}_{0n}}{2\bar{N}_{11}} -iy_1 e^K S_{0n}\right)=0,\nonumber \\
I&=&2:\qquad\frac{y_2 e^{K/2}}{2}\pa_n \ln \frac{y_1 y_3}{ y_2 } +4e^{-U}\left(     
 -iy_2 e^K S_{0n}\right)=0,\nonumber \\
I&=&3:\qquad\frac{y_3 e^{K/2}}{2}\pa_n \ln \frac{y_1 y_2 }{y_3}  +4e^{-U}\left(           
 -iy_3 e^K S_{0n}\right)=0 .
\ea
Here using (\ref{M5}) and (\ref{E16}) we expressed the field strengths in the
 section with the prepotential through the strengths in the
heterotic section.

The system of equations (\ref{E30}) and (\ref{F3}) is solved by a general
 magnetic black hole with two arbitrary magnetic charges. The background
consist of a metric, dilaton and moduli \cite{cvyo,bh2}. In the
following, we shall consider a particular extremal solution 
\ba
\label{E32}
e^{-2U} = 1+\frac{P}{r}, \quad y_1 =e^{-\p}=\left(1+\frac{P}{r}\right)^{-1}, 
\ea
The charges $P^0$ and $P^1$ are expressed via a charge $P$ 
\be
\label{E33}
 P^0 = \frac{P^1 }{y_2 y_3 },\quad P= \sqrt{8 y_2 y_3}P^0
\ee
and the moduli $y_2,\,y_3$ are  real constants. The metric
components of the torus $T^2$ are
$$ G_{11}=y_2 y_3 =e^{2\g}, \qquad  G_{22}=y_2 /y_3 = e^{2\s}.
$$ 
 As in (\ref{E6}), the factor $\sqrt{8}$ appears because of different 
normalizations of the gauge terms in the actions (\ref{E1}) and (\ref{F1}).
\section{Solution of the loop-corrected spinor Killing equations}
Our next aim is to solve the Maxwell equations  and the loop-corrected
spinor Killing equations.
 We look for a solution in the
first order in  string coupling constant. The loop
corrections are calculated by substituting the tree-level moduli.
For the constant  moduli $y_2,\,y_3$, the loop correction to the
prepotential and its derivatives are  also independent of coordinates, 
resulting in considerable technical simplifications in solution 
of the loop-corrected spinor Killing equations.

In the holomorphic section associated with the heterotic string
compactification, the Maxwell equations (\ref{E12}) 
which we rewrite as
\be
\label{E35}  
\pd (\sqrt{-g}\, Im \hn_{IJ} \h{\F}^J +Re \hn_{IJ}\z \h{\F}^J )^{\m\n} =0,
\ee
with the required accuracy have the form
\ba
J=0&:&\quad \pa_r [\sqrt{-g} \,Im \hn_{00} \h{\F}^0 +
Re\hn_{00}\z\h{\F}^0 + Re\hn_{01}\z\h{\F}^1 ]^{0r} =0, \label{E36}\\
J=1&:&\quad \pa_r [\sqrt{-g}\, Im \hn_{11} \h{\F}^1  + 
Re \hn_{10}\z\h{\F}^0 +Re\hn_{11}\z\h{\F}^1]^{0r} =0, \label{E37}\\
J=2&:&\quad \pa_r [\sqrt{-g}\,  Im \hn_{22} \h{\F}^2  + 
Re \hn_{20}\z\h{\F}^0 +
 Re \hn_{21}\z\h{\F}^1  ]^{0r} =0, \label{E38}\\
J=2&:&\quad \pa_r [\sqrt{-g}\,  Im \hn_{33} \h{\F}^3  + 
Re \hn_{30}\z\h{\F}^0 + Re
\hn_{31}\z\h{\F}^1  ]^{0r} =0. \label{E39}
\ea
Here $\z \F^{0r} =\F_{\vt\vp} $ and $\z \F_{\vt\vp}=-\F^{0r}$. 
Only the diagonal gauge couplings $\hn_{II}$ contain  terms of
zero-order in  string coupling. 
The tree-level gauge
field strengths acquire corrections of the first order in  string
coupling, and also the gauge fields, absent at the tree level,
are generated. 

First, we solve the equations (\ref{E38}) and (\ref{E39}):
\ba 
\label{E40}
\h{\F}^{2\,0r} =\frac{ C_2 - (Re N_{20}+c_{20})P^0 - Re
\frac{N_{21}}{N_{11}} P^1}{\sqrt{-g'}\, Im \hn_{22}} 
,\nonumber \\
\h{\F}^{3\,0r} =\frac{ C_3 -( Re N_{30}+c_{30})P^0 - Re
\frac{N_{31}}{N_{11}} P^1}{\sqrt{-g'}\, Im \hn_{33}} . 
\ea

 Eqs. (\ref{E36}) and (\ref{E37}) yield
\ba 
\label{E41}
\h{\F}^{0\,0r} &=&\frac{C_0 -c_{00} P^0 -a_1 P^1 }{\sqrt{-g'}\,Im \hn_{00}},\nonumber \\
\h{\F}^{1\,0r} &=&\frac{C_1 -d_{11} P^1 -a_1 P^0 }{\sqrt{-g'}\,Im 
\hn_{11}}
\ea
where $\sqrt{-g'} = e^{-2U}r^2$ and $C_I$ are arbitrary constants of
the first order in  string coupling.  All the electric fields $\h{\F}^{I}$
are of the first order in string coupling.

One can also introduce  magnetic fields $\z\F^2$ and $\z\F^3$ with the
charges of the first order in string coupling. However, since these fields
enter the Maxwell equations multiplied by the coupling constants of the
first order in string coupling, at the required level of accuracy, these
terms  are omitted from the equations. By the same reason the terms with the
constants $c_{22}$ and  $c_{33}$ do not appear in solutions also.
  
Let us calculate the symplectic invariant combination $S_{0n}$ (\ref{E18})
in the first order in  string coupling constant. 
We have
\ba
\label{E43}
S_{0n}&=&\left\{ P^0 [Im\, N_{00} +y_i Re N_{i0} 
 +i(a_1 y_2 y_3 +a_2 y_1 y_3 + a_3 y_1 y_2 )]  
-P^1 y_1 \right. \nonumber  \\&-& \left.
[ (y_2 C_2 +y_3 C_3) +i(C_0 -c_{00}\,P^0 +
(C_1 -d_{11}\, P^1 ) y_2 y_3 ) ]\right \}\frac{i}{2}e^{2U}\frac{x^n }{r^3 } .
\ea
Only the the couplings $ N_{00}$ and $N_{0i}, \, i=1,2,3$ enter the final 
expression (\ref{E43}) yielding
$$Im\, N_{00} +y_i Re N_{i0} = -(y_1 y_2 y_3  +2v +h_a y_a ).
$$
All the terms containing second derivatives of the
prepotential have canceled. 

Since gaugino spinor Killing equations are linear in derivatives of the moduli, and
axions $a_i$ are of the first order in string coupling, the equations for
real parts of the moduli 
decouple from the equations for axions. In the first order in string
coupling, the K\"{a}hler potential (\ref{E10}) is also
independent of the axions $a_i $.
The  constants $C_0$ and $C_1$
enter only the imaginary parts of spinor Killing equations. In the following
we shall discuss only the real parts of spinor Killing equations, i.e.
equations for the real parts of the moduli. The
equations for the axions will be considered elsewhere.
  
Using the K\"{a}hler potential (\ref{E10}), we calculate  the combinations 
$B_n^I =f^I_i \pa_n z^i$ which enter the
spinor Killing equations (\ref{E26}) for the moduli $z_i$. 
We have 
\ba
\label{E44}
B_n^0 &=& -{1\over 2}e^{K/2 } \left(1-{V\over 2y_1 }\right)\pa_n\ln y^3 \nonumber \\
B_n^i &=&iy_i \left(B_n^0 +e^{K/2}\pa_n\ln y_i \right), \qquad i=1,2,3. 
\ea
Here we substituted
$$ V=e^{-2\g}v.
$$
 All the expressions are calculated in the first
order in  string coupling. In particular, all the factors multiplying the
Green-Schwarz function $V$ are taken in the leading order in string
coupling. 
 
Let us introduce the notations for the loop-corrected metric and moduli.
We shall split  the functions
$\p, \g $ and $\s$ which appear in the moduli (\ref{E3}) into the tree-level
parts $\p_0 , \g_0 $ and $\s_0$ and the those of the first order in string
coupling: $\p_1 , \g_1 $ and $\s_1$. We have  $\p =\p_0 +\p_1$, etc..
 The function $2U$ in the metric
will be written as $2U_0 +u_1 $. At the tree level (see (\ref{E32}),
$$ e^{-2U_0 } = e^{\p_0 } =\f0 = 1+{P\over r} 
$$
From (\ref{E33}), we have
\be
\label{E45}
P^0 = \frac{Pe^{-\g_0}}{\sqrt{8}},\qquad P^1 = \frac{Pe^{\g_0}}{\sqrt{8}}
\ee
For the K\"{a}hler potential we obtain
\be
\label{E46}
e^K = \frac{\f0 e^{-2\g_0 }}{8}\left[1 +\left(\p_1 -2\g_1 -{V\f0\over
2}\right)\right] .  
\ee

Let us turn to the function $S_{\vt\vp}$ (\ref{E43}).
 The functions  $a_i$ are of
the first order in the string coupling constant.  The terms
containing the factors $a_i$ are imaginary. Because  the spinor Killing
equations for the moduli (\ref{E25}) 
 are linear in derivatives of the moduli, the equations for the
imaginary parts of the moduli decouple from those for the real parts. In
this section, solving the equations for the real parts of the moduli, 
 the imaginary terms in the equations will be omitted.

Because the tree-level moduli are constants, the terms $2v + h_a y_a$ and 
$y_a C_a$, which are of the first  order in string coupling, are also 
constants.

The function $T^-_{0n}$  (\ref{E18}) which enters the gravitini equation 
(\ref{E23}), we write in the form
\be   
\label{E47}
T^-_{0n}=e^{K/2} y_1 \left[P^0 e^{2\g }\left(1+\left(2V+h_a y_a e^{-2\g_0
}\right)\f0\right) +P^1 +C_a y_a  \f0 \right]e^{2U}\frac{x^n}{r^3}.
\ee
  The factors $\f0$ appear because in the expressions of the first order in
string coupling the modulus 
$y_1$  can be substituted by its tree-level value ${\f0}^{-1}$.

Using the expression for the K\"{a}hler potential (\ref{E46}) and expanding in
(\ref{E47}) all the terms to the first order in string coupling, we finally
obtain
\be   
\label{E48}
T^-_{0n} = \f0^{-1/2} \frac{P}{4} \left[1 + \left(-\frac{\p_1}{2} + u_1 +
\left(\frac{3V}{4}+C\right)\f0 \right)\right]\frac{x^n}{r^3},
\ee
where the constant $C$ is
\be
\label{F5}
C={1\over2}\left(h_a y_a  +\frac{C_a y_a }{P^0 }\right)e^{-2\g_0 }.
\ee
Substituting the expression (\ref{E48}), we obtain the gravitino spinor 
Killing equation in the form
\be   
\label{E49}
\frac{1}{4}\pa_n [\f0^{-1}(1+u_1 )] -\frac{P}{4}\f0^{-2}\left[1+\left({3u_1 \over 2}-
{\p_1 \over 2}+\left(\frac{3V}{4}+C\right) \right)\right]\frac{x^n}{r^3} =0.
\ee
In this equation the combination of the tree-level terms vanishes; 
the remaining part of the first order in string coupling is
\be   
\label{E50}
\frac{u'_1}{q'} +\frac{u_1 -\p_1}{2} +\left(\frac{3V}{4}+C
\right){\f0} =0.
\ee
Here $q'= {\f0}'/\f0$.

Let us turn to the gaugini spinor Killing equations (\ref{E26}).
Substituting  the expression
\be
\label{F7}
e^K S_{0n}=-{1\over4}\left(1-\g_1
+\left(\frac{V}{2}+C\right)\f0 \right)\left(\frac{Pe^{-\g_0 }}{\sqrt{8}}
{i\over2}e^{2U}\frac{x^n}{r^3}\right),
\ee
for the combination
$\frac{1}{2} \F^{-0}_{0n} +e^K S_{0n}$ 
we have
\be   
\label{E51}
\frac{1}{2} \F^{-0}_{0n} +e^K S_{0n}={1\over4}\left(1+\g_1
-\left(\frac{V}{2}+C\right)\f0 \right)\left(\frac{Pe^{-\g_0 }}{\sqrt{8}}
{i\over2}e^{2U}\frac{x^n}{r^3}\right).
\ee
For the combination $\frac{1}{2} \F^{-1}_{0n} - iy_1 e^K S_{0n}$ we obtain 
\be   
\label{E53}
\frac{1}{2} \F^{-1}_{0n} -iy_1 e^K S_{0n}=-\frac{iy_1}{4}\left(1-3\g_1
 +\left({V\over2}+C\right)\f0\right)\left(\frac{Pe^{-\g_0 }}{\sqrt{8}}
{i\over2}e^{2U}\frac{x^n}{r^3}\right).
\ee

With the accuracy of the terms of the first order in string coupling, the  
loop-corrected expressions for  ${B^0}_n$ and ${B^1}_n$ are
\ba   
\label{E54}
{B^0}_n =\frac{ q' {\f0}^{1/2 }e^{-\g_0 } }{2\sqrt{8}}
\left[1+\frac{{\p_1 }'-2{\g_1 }'}{q'}+
\frac{\p_1 -2\g_1 }{2} -\frac{3V\f0 }{4}\right]\frac{x^n}{r},
\nonumber \\
{B^1}_n =-i\frac{q' \f0^{-1/2}e^{-\g_0}}{2\sqrt{8}}
\left[1+\frac{{\p_1}' +2{\g_1}' }{q' }-\frac{\p_1 +2\g_1 }{2} +
\frac{V\f0 }{4}\right]\frac{x^n}{r}.
\ea

Using the expressions (\ref{E51})-(\ref{E54}), we verify that in the gaugini
spinor Killing  Eqs.(\ref{E26})  with $I=0$ and $I=1$  the leading-order
terms cancel, and the remaining parts of the 
first order in string coupling are
\ba
\label{E55}
I=0:\qquad\frac{{\p_1}'-2{\g_1}' }{q' } +\frac{\p_1 -u_1 }{2} -2\g_1
-\left({V\over4}-C \right)\f0 =0, \nonumber \\
I=1:\qquad\frac{{\p_1}'+2{\g_1}' }{q' } +\frac{\p_1 -u_1 }{2} +2\g_1
-\left({V\over4}-C \right)\f0 =0.
\ea
 Eqs. (\ref{E55}) split into the following system
\ba
\label{E56}
\frac{{\p_1}' }{q' } +\frac{\p_1 -u_1 }{2} 
-\left({V\over4}-C \right)\f0 =0 \nonumber \\
{\g_1 }' +q' \g_1 =0
\ea 

Let us consider the  gaugino spinor Killing equations equations  
(\ref{E26}) with $I=2$ and $I=3$.
Substituting in the expressions for  the loop-corrected
couplings (\ref{E34}) and the field strengths (\ref{E40}), we have
\be
\label{E57}
\F^{-2}_{0n} =\frac{P^0 }{y_1 y_3 } \left({v\over2 }+h_2 y_2  +\frac
{C_2 y_2 }{P^0 } \right) \frac{1}{2} e^{2U}\frac{x^n}{r^3} = 
y_2 P^0 \f0 \left({V\over2}+L_2 \right)\frac{1}{2} e^{2U}\frac{x^n}{r^3}  
\ee
 and similar expression for $\F^{-3}_{0n} $ obtained by substitution
$2\rightarrow 3$.  
The field strengths $\F^{-2,3}$, absent at the string tree level, 
are of the first order in the string
coupling. Here we introduced
\be
\label{E58}
L_2 =\left(h_2 y_2 +\frac{C_2 y_2 }{P^0 }\right)e^{-2\g_0 }, \qquad
L_3 =\left(h_3 y_3 +\frac{C_3 y_3 }{P^0 }\right)e^{-2\g_0 }. 
\ee

Subtracting  Eq.(\ref{E34}) with $I=2$ from that with $I=0$ (the same
for $I=3$ ), and using the expressions (\ref{E44}) for the combinations $B^i_n $, we have
\be
\label{E59}
ie^{K/2}\frac{\pa_n y_2 }{y_2 } +4e^{-U}\left(\frac{\h{\F}^{-2}_{0n}}{2iy_2 }
-\frac{1}{2}\h{\F}^{-0}_{0n} -2e^K S_{0n}\right) =0.
\ee
Substituting the expressions for the field strengths $\h{\F}^{-0}_{0n}$,
$\h{\F}^{-2}_{0n}$ and Eq.(\ref{F7}) for $e^K S_{0n}$, and keeping the terms of the
first order in the string coupling, we obtain
\ba
\label{E60}
{\g_1}' +{\s_1}' +(C-L_2 -\g_1 \f0^{-1})\frac{P}{r^2}=0\nonumber \\
{\g_1}' -{\s_1}' +(C-L_3 -\g_1 \f0^{-1})\frac{P}{r^2}=0   .
\ea
The sum of the Eqs. (\ref{E60}) is
\be
\label{E61}
{\g'}_1 +\g_1 q' +(2C-L_2 -L_3 )\f0' =0.
\ee
Substituting the expressions (\ref{F5}) and (\ref{E58}) for $C$ and $L_a$, we find 
that
\be
\label{E62}
2C-L_2 -L_3 =0,
\ee 
so that Eq.(\ref{E61}) coincides with the second Eq.(\ref{E56}).

Let us solve the system of equations for the functions $u_1$ and $\p_1$. 
 Adding and subtracting the gravitini  Eq.(\ref{E50} and the
first Eq.(\ref{E56}), we obtain the solution
\ba 
\label{E63}
u_1 +\p_1 & =&c_1 -\left(\frac{V }{2}+2C\right)\f0 \nonumber \\
u_1 -\p_1 &=&\frac{c_2}{\f0 } -\frac{V\f0 }{2},
\ea
where $c_{1,2}$ are arbitrary constants. Requiring that at large distances
from the center of the black hole the metric and dilaton are asymptotic to
the Lorentzian metric and constant dilaton equal to unity, we have
\be
\label{E64}
c_1 =\frac{V}{2} +2C, \qquad c_2 =\frac{V}{2},
\ee
and we obtain
\be                                                  		
\label{E65}                                          		
u_1 = -\left(\frac{V }{2}+C \right)\frac{P}{r}-\frac{V }{4}\frac{P}{r+P},
\qquad
\p_1 = -C{P\over r}+\frac{V }{4}\frac{P}{r+P}.
\ee

At the tree level, magnetic black hole solution is the extremal BPS
saturated configuration \cite{cvyo,bh2}. Provided supersymmetry is 
unbroken in perturbation theory, the loop-corrected solution  must have
the same properties. 
ADM mass can be obtained from the $r\rightarrow\infty$ asymptotics of the
metric. Using (\ref{E65}) we have
$$ M_{ADM}=2P(1-\frac{3V}{4}-C).
$$
BPS mass is determined from the asymptotics of the function $T^-$ which can
be written as 
$$ T^-_{\m\n} = (F_I \F^I_{\m\n} -X^I G_{I\m\n} ).
$$
Asymptotics of the fields $\F^I$ and $G_I$ are
proportional to  electric and magnetic charges, correspondingly, and
the asymptotics of  $T^-$ yields the expression for the BPS mass \cite{cafpr}
 $M_{BPS} =|Z_{\infty}| = e^{K/2} |n_I X^I - m^I F_I |_{\infty}$.
Taking the asymptotics of the of the function $T^-$ (\ref{E48}), we find
the BPS mass  which is equal to the ADM mass.
 
A particular possibility 
 is to take $C=-\frac{3V}{4}$, in which case  the central charge 
 retains the tree-level form. Correction to the metric becomes
\be                                                             
\label{E66} 
u_1 =\frac{V}{4}\left({P\over r} -\frac{P}{r+P} \right).
\ee  
With required accuracy, the loop-corrected metric can be considered
as the leading and the
first-order terms in the expansion of the expression
\be
\label{E67}  
 g_{ii}=-g^{00} = 1+\frac{P}{r +\e\frac{V}{4}}  
\ee
in powers of the string-loop counting parameter $\e$. In the metric
(\ref{E67}) the singularity at the point $r=0$ is smeared by the loop
correction.
\section{Discussion}
In our previous study \cite{mi}, solving the system of 
the loop-corrected Einstein and
Maxwell equations, we obtained a two-parameter set of
solutions for the loop corrections to the metric and dilaton
\be                                                             
\label{E68}
u_1 =A_1 \frac{P}{r}  - A_2\frac{P}{r+P}, \qquad
\p_1 =(A_1 +\frac{V}{2})\frac{P}{r}+A_2\frac{P}{r+P}.
\ee
The one-parameter family of solutions (\ref{E65}) is contained in the set
(\ref{E68}). A nontrivial check of consistency of both calculations 
is that in both cases the coefficients 
at the terms $\frac{P}{r}$ in the expressions for the metric and dilaton 
differ by $\frac{V}{2}$.

Near the locations of the enhanced symmetry points in the moduli space, 
the second derivatives of the
prepotential have logarithmic singularities \cite{afgnt,wikalu}. In particular, 
for $y_2 \sim y_3 $, 
$$h^{(1)}(y_2 ,y_3 )=(y_2 -y_3 )^2 \log(y_2 -y_3 )^2 .
$$
Although the loop-corrected gauge couplings (\ref{E34}) contain second
derivatives of the prepotential, the final expressions for the 
metric and moduli depend on
the Green-Schwarz function $V$ which contains only the first derivatives
of the prepotential  and thus is regular at the points of enhanced symmetry. 
Note also, that the Green-Schwarz function is positive \cite{kou} (this can be verified
by explicit calculations) as can be seen from the form of the K\"{a}hler
potential for the moduli which is a regular function at finite values of the
moduli.

Our solution for the loop corrections is valid for all $r$ for which is
valid the perturbation expansion in string coupling. In particular, since
the dilaton increases at small distances, we
can use both the tree-level and a loop-corrected solution for  
 $\frac{r}{P}>\e V$.  However, if we extrapolate the above expression for
the loop-corrected  metric  to the
region of small $r$, it can be seen that singularity at the origin is
smeared. The crucial point is that the Green-Schwarz function $V$ is
positive \cite{kou}.

Our treatment of spinor Killing equations is similar in spirit to
\cite{fre,bt}. However, in these papers were discussed only tree-level spinor
Killing equations. Another distinction is that usually
 an emphasis was
made on the form of solution  at the stabilization
point\cite {kal}, whereas we were
interested in full coordinate dependence of solution.

Our approach is different from that in papers \cite{bgl} based on the
assumption that there is a "small" modulus which can be used as an expansion
parameter for the loop-corrected action. In string-loop perturbative
expansion, a natural expansion parameter is associated with the dilaton, and
the loop correction to the tree-level prepotential is independent of the
modulus $y_1\equiv S$.

Finally, in  perturbative approach, we neglected the terms of the
form $O(e^{2\pi S})$, and the duality properties of the full theory
\cite{nper} cannot be checked in this approximation.
\bigskip
\bigskip

{\large \bf Acknowledgements}

\medskip

\noindent
I thank Renata Kallosh for helpful correspondence and M. Bertolini and
M. Trigiante for bringing to my attention reference \cite{bt}.
This work was  partially supported by the RFFR  grant No 00-02-17679.

\end{document}